%% file: mdt.tex
\documentclass[submission,copyright,creativecommons]{eptcs}
\providecommand{\event}{FPCS 2015} 
\usepackage{breakurl}             

\title{Reasoning about modular datatypes with Mendler induction}
\author{ Paolo Torrini \qquad \qquad Tom Schrijvers
  \institute{\\Department of Computer Science, KU Leuven, Belgium} 
\email{\quad \{p.torrini, tom.schrijvers\}@cs.kuleuven.be}
}
\def\titlerunning{Modular data-types with Mendler induction}
\def\authorrunning{P. Torrini \& T. Schrijvers}

\usepackage{mathtools} 
\usepackage{amssymb}
\usepackage{latexsym}
\usepackage{stmaryrd}
\usepackage{ifthen}

\usepackage{amsmath}
\usepackage{hyperref}

\usepackage{proof}
\usepackage{comment}

\include{lightmacros2}

\begin{document}
\maketitle              

\begin{abstract}
In functional programming, datatypes \`{a} la carte provide a
convenient modular representation of recursive datatypes, based on
their initial algebra semantics. Unfortunately it is highly
challenging to implement this technique in proof assistants that are
based on type theory, like Coq. The reason is that it involves type
definitions, such as those of type-level fixpoint operators, that are
not strictly positive. The known work-around of impredicative
encodings is problematic, insofar as it impedes conventional inductive
reasoning.  Weak induction principles can be used instead, but they
considerably complicate proofs.

This paper proposes a novel and simpler technique to reason
inductively about impredicative encodings, based on Mendler-style
induction. This technique involves dispensing with dependent
induction, ensuring that datatypes can be lifted to predicates and
relying on relational formulations. A case study on proving subject
reduction for structural operational semantics illustrates that the
approach enables modular proofs, and that these proofs are essentially
similar to conventional ones.
\end{abstract}
%


\section{Introduction} 
\label{Introduction}

Developing high-quality software artifacts, including programs as well
as programming languages, can be very expensive, and so can formally
proving their properties. This makes it highly desirable to maximise
reuse and extensibility. Modularity plays an essential role in this
context: a component is modular whenever it can be specified
independently of the whole collection -- therefore, a modular
characterisation of an artifact implies that its extension does not
require changes to what is already in stock.

In functional programming, it is natural to rely on a structured
characterisation of components based on recursive datatypes. However,
conventional datatypes are not extensible -- each one fixes a closed
set of constructors with respect to which case analysis may have to be
exhaustive, hence each case implicitly depends on the whole
collection. An elegant solution to this tension between structural
characterisation and modularity, also known as the \emph{expression
  problem}, has been found with the notion of \emph{modular datatype}
(MDT) -- i.e., datatypes \emph{\`{a} la carte}, introduced in Haskell
by Swierstra \cite{Swier08}.  The definition of an MDT consists of two
distinct parts: the grammar, as a non-recursive structure based on a
functor, and the recursive datatype, as the recursive closure of the
functor by a type-level fixed point. Grammar functors behave as
modules, as they can be defined independently and combined together by
coproduct.

In Haskell, an MDT can be easily implemented in terms of conventional
datatypes, which can be used to define the grammar as well as the
recursive closure (as recalled in
Section~\ref{section:zero}). However, Haskell's datatype definition of
the type-level fixpoint operator is not strictly positive, and
therefore it is problematic from the point of view of less liberal
type systems.
As a general-purpose programming language, Haskell relies on types
that do not enforce totality (i.e., either termination or
productivity). This makes type checking easier in the presence of
non-termination. Unfortunately, allowing for non-total programs can
lead to inconsistency under a program-as-proof interpretation. For
this reason, proof assistants based on the Curry-Howard correspondence
are usually based on more restrictive type systems.
Proof assistants such as Coq, Agda, Isabelle and Twelf, for instance,
rely on a syntactic criterion of monotonicity which ensures totality,
by requiring that all the occurrences of an inductive datatype in its
definition are strictly positive -- hence incompatibly with the
Haskell-style representation of MDTs. 

Coq is a theorem prover based on the calculus of inductive
constructions (CIC) \cite{coqart} which extends the calculus of
constructions (CC) \cite{COQ} with inductive and coinductive
definitions. CC, the most expressive system of the lambda cube
\cite{Bar92}, allows for types depending on terms, type-level
functions and full parametric polymorphism, hence also for definitions
that are impredicative, in the sense of referring in their bodies to
collections that are being defined.
One of the main approaches to represent MDT in Coq, due to Delaware,
Oliveira and Schrijvers \cite{Dela13} and implemented in the MTC/3MT
framework \cite{Delaware13M}, takes advantage of impredicativity, and
relies on the Church encoding of fixed points (as recalled in
Section~\ref{section:one}). Another promising approach, due to Keuchel
and Schrijvers \cite{Keuchel13}, relies on containers -- it is
predicative, but it involves a more indirect representation of
types. Church encodings are purely based on CC and do not involve any
extra-logical machinery -- however, they rather complicate inductive
reasoning.  Impredicative definitions have an eliminative character
that hides term structure, hence making it harder to reason by
induction. The solution proposed by Delaware \emph{et al.} is quite
general -- however, it relies on proof algebras that pack terms
together with proofs using $\Sigma$-types, and this leads to inductive
proofs that have a significant overhead with respect to the
conventional, non-modular ones.


This paper proposes a novel solution to the problem of reasoning
inductively with impredicatively encoded MDT, based on the use of
Mendler-style induction
\cite{Mendler91,UustaluV99,AbelMU05}. Mendler's characterisation of
iteration makes it possible to encode an induction principle within
the impredicative encoding of an MDT. Unlike Delaware \emph{et al.},
we use Mendler algebras as proof algebras. This leads to inductive
proofs that are straightforwardly modular and ultimately closer to
conventional ones (Section~\ref{section:two}). Although this approach
cannot handle dependent induction, this limitation is of little
consequence as long as we are reasoning about relational
formulations. Nonetheless, this may make it necessary to lift
inductive datatypes to inductively defined predicates, in order to use
them as inductive arguments in proofs.


In order to reason inductively on relations, we clearly need to rely
on functor shapes that can represent them as well as mutual
dependencies. Such need is highlighted throughout a case study on the
formalisation of a language based on structural operational semantics
(Section~\ref{section:three}, Coq implementation available
\cite{MACoq}). The language, for which we prove type preservation, has
a definition that involves mutual dependency between expressions and
declarations.

\section{Datatypes a-la-carte}  \label{section:zero}

MDTs as introduced by Swierstra \cite{Swier08} are essentially a
functional programming application of the initial algebra semantics of
inductive types. This consists of associating an inductive datatype to
an endofunctor in a base category, then interpreting it as the initial
object in the category of algebras determined by the functor
\cite{Hagino87,WadlerRecTypes}.

In its simplest form, taking sets ($\msf{S}$) as the base category,
each inductive datatype $\varrho : \msf{S}$ can be associated with a
covariant endofunctor (\emph{signature functor}), i.e. a map $F :
\msf{S} \to \msf{S}$ for which there exists a map (\emph{functor map})
$\msf{fmap}_F \ \{ A \ B \}: \ (A \to B) \to (F \ A \to F \ B)$ that
preserves identities and composition, with $A, \ B: \msf{S}$ (always
treated as implicit parameters). Semantically, an algebra determined
by $F$ ($F$-\emph{algebra}) is a pair $\langle C, \phi \rangle$ where
$C:\msf{S}$ is the \emph{carrier} and $\phi: F \ C \to C$ is the
\emph{structure map}. $F \ C$ can be understood as the denotation of a
grammar based on signature $F$, given carrier $C$. The initial object
$\langle \mu F, \msf{in}_F \rangle$, where $\msf{in}_F$ is an
isomorphism and thus has an inverse $\msf{out}_F$, gives the
denotation of $\varrho$ obtained as the fixpoint closure of $F$.
In this way, the non-recursive structural characterisation of
$\varrho$, which essentially corresponds to case analysis, is
separated from its recursive closure. For instance, in a functional
language which allows for datatype definitions with data constructors
and Haskell-style destructors (while we mainly rely on Coq-style and
standard algebraic notation), the following
\begin{gather} \label{data-rho}
 \msf{dt\_def} \ \varrho \ = \ \msf{c}_1 \ (\tau_1[\varrho/A]) \mid
\ldots \mid \msf{c}_k \ (\tau_k[\varrho/A]) 
\end{gather}
can be decomposed in
\begin{gather}
\msf{dt\_def} \ F \ A \ = \ \msf{c}_1 \ (\tau_1) \mid \ldots \mid
\msf{c}_k \ (\tau_k)
\end{gather}
and
\begin{gather}
 \varrho \ \defeq \ \msf{Fix} \ F 
\end{gather}
where $\msf{Fix} \ F$ is the syntactic representation of $\mu F$, i.e.
\begin{gather}
 \msf{dt\_def} \ \msf{Fix} \ F \ = \ \msf{in} \ (\msf{out} : \ F
 \ (\msf{Fix} \ F))
\end{gather}
For each $F$-algebra $\langle C, f \rangle$, the unique incoming
algebra morphism from the initial algebra is determined by the unique
\emph{mediating map} $ \msf{fold}_{F,C,f} : \mu F \to
C$. Syntactically, this corresponds to the definition of $\msf{fold}
\ F \ C : \ (F \ C \to C ) \to (\msf{Fix} \ F \to C)$ as a recursive
function.
%
\begin{gather}
 \msf{fold} \ F \ C \ f \ x \ \defeq \ f \ (\msf{fmap} \ F
 \ (\msf{fold} \ F \ C \ f) \ (\msf{out} \ x))
\end{gather}
%
Functors are composable by coproduct ($\mbox{+}$), i.e., if $F_1, F_2:
\msf{S} \to \msf{S}$ are functors, so is $F_1 \mbox{+} F_2$, with
\begin{gather}
 \msf{dt\_def} \ (F_1 \mbox{+} F_2) \ C \ = \ \msf{inl} \ (F_1 \ C)
 \mid \msf{inr} \ (F_2 \ C)
\end{gather}
This results in a modular definition of the inductive datatype
$\msf{Fix} \ (F_1 \mbox{+} F_2)$ -- not to be confused with $\msf{Fix}
\ F_1 \mbox{+} \msf{Fix} \ F_2$.  In connection with coproducts,
Haskell implementations of MDTs rely on type classes to automate
injections and projections, using smart constructors and class
constraints to express subsumption between functors.  As a concrete
example, following Swierstra \cite{Swier08}, the conventional datatype
\begin{gather} \label{data-rhoX1}
 \msf{dt\_def} \ \msf{Trm} \ = \ \msf{lit} \ (\msf{Int}) 
  \mid \msf{add} \ (\msf{Trm} * \msf{Trm})
\end{gather}
can be decomposed into two modules
\begin{gather} \label{data-rhoX2}
 \msf{dt\_def} \ \msf{Trm_{G1}} \ C \ = \ \msf{lit} \ (\msf{Int})
 \qquad \qquad
 \msf{dt\_def} \ \msf{Trm_{G2}} \ C \ = \ \msf{add} \ (C * C)
\end{gather}
and thus modularly defined: 
\begin{gather} \label{data-rhoX3}
\msf{Trm_G} \ =_{df} \ \msf{Trm_{G1}} + \msf{Trm_{G2}} \qquad \qquad 
 \msf{Trm} \ =_{df} \ \msf{Fix} \ \msf{Trm_G}
\end{gather}
Moreover, given a notion of value and a conventional recursive
definition of evaluation
%
%
\begin{gather} \label{data-rhoX5}
\begin{array}{lll}
 \msf{dt\_def} \ \msf{Val} \ = \ \msf{val} \ (\msf{vv}: \msf{Int})
&
\qquad \msf{eval} : \msf{Trm} \to \msf{Val} \\
& \quad \msf{eval} \ (\msf{lit} \ x) \ =_{df} \ \msf{val} \ x \\
& \quad \msf{eval} \ (\msf{add} \ (e_1,e_2)) \ =_{df} \ \msf{val}
\ ((\msf{vv} \circ \msf{eval} \ e_1) + (\msf{vv} \circ \msf{eval}
\ e_2))
\end{array}
\end{gather}
the latter can be represented by an algebra and modularly decomposed
as follows, allowing for a modular definition of the dynamic
semantics.
\begin{gather} \label{data-rhoX6}
\begin{array}{lll}
\msf{eval_{G1}} : \msf{Trm_{G1}} \ \msf{Val} \to \msf{Val} \qquad \qquad 
\msf{eval_{G1}} \ (\msf{lit} \ x) \ =_{df} \ \msf{val} \ x  \\
%
\msf{eval_{G2}} : \msf{Trm_{G2}} \ \msf{Val} \to \msf{Val}  \qquad \qquad
\msf{eval_{G2}} \ (\msf{add} \ (x_1, x_2))
\ =_{df} \ \msf{val} \ ((\msf{vv} \ x_1) + (\msf{vv} \ x_2)) \\
%
\msf{eval_G} : \msf{Trm_G} \ \msf{Val} \to \msf{Val}  \qquad \qquad \ 
\msf{eval_G} \ (\msf{inl} \ e) \ =_{df} \ \msf{eval_{G1}} \ e \\
\qquad \qquad \qquad \qquad \qquad \qquad \qquad \msf{eval_G} \ (\msf{inr}
\ e) \ =_{df} \ \msf{eval_{G2}} \ e
\end{array} \\
\label{data-rhoX7}
\msf{eval} \ e \ =_{df} \ 
   \msf{fold} \ \msf{Trm_G} \ \msf{Val} \ \msf{eval_G} \ e
\end{gather}
%
%
%

\section{Impredicative encoding}  \label{section:one}

The MDT representation discussed so far works well with Haskell, but
not with Coq. Representing $F$ as an inductive datatype is not
problematic, but this is not so for the fixpoint closure. Since the
constructor of $\msf{Fix} \ F$ has type $ F \ (\msf{Fix \ F}) \to
\msf{Fix} \ F$, the datatype has a non-strictly positive occurrence in
its definition, as parameter of the argument type -- hence it is
rejected by Coq. There is an analogous issue with the definition of
$\msf{fold}$, which is not structurally recursive. The solution to
this problem adopted by Delaware \emph{et al.} in \cite{Dela13}, which
we summarise here, goes back to Pfenning and Paulin-Mohring
\cite{PfenningP89} in relying on a Church-style encoding of fixpoint
operators, thus requiring impredicative definitions.


From the point of view of a type theoretic representation, the type of
an algebra (that we may call \emph{Church algebra}, or conventional
algebra) can be identified with the type of its structure map.
\begin{gather}
 \msf{Alg^C} \ F \ C \ \defeq \ F \ C \to C 
\end{gather}
If the initiality property of fixed points is weakened to an existence
property, a fixpoint operator can be regarded as a function that maps
an algebra to its carrier. An abstract definition of the type-level
fixpoint operator $\msf{Fix^C}: (\msf{S} \to \msf{S}) \to \msf{S}$ can
then be given, as elimination rule for $F$-algebras, impredicatively
with respect to $\msf{S}$ (this requires the impredicative set option
in Coq, as used in MTC/3MT \cite{Dela13}).
\begin{gather}
 \msf{Fix^C} \ F \defeq \forall A: \msf{S}. \ \msf{Alg^C} \ F \ A \to A
\end{gather}
%
The map $\msf{fold^C} \ F \ C : \msf{Alg^C} \ F \ C \to \msf{Fix^C}
\ F \to C$, corresponding to the elimination of a fixpoint value, can
now be defined as the application of that value.
\begin{gather}
\msf{fold^C} \ F \ C \ f \ x \defeq x \ C \ f
\end{gather}
Relying on the functoriality of $F$, the in-map $\msf{in^C} \ F : F
(\msf{Fix} \ F) \to \msf{Fix} \ F$ and the out-map $\msf{out^C} \ F :
\msf{Fix} \ F \to F (\msf{Fix} \ F)$ can be defined as functions.
\begin{gather}
\msf{in^C} \ F \ \defeq \ \lambda x \ A \ f. \ f (\msf{fmap} \ F
\ (\msf{fold^C} \ F \ A \ f) \ x) \\
%
 \msf{out^C} \ F \ \defeq \ \msf{fold^C} \ F \ (F (\msf{Fix} \ F))
\ (\msf{fmap} \ F \ (\msf{in^C} \ F)) 
\end{gather}
Notice that the definition of $\msf{fold^C} \ F \ C \ f$ does not
guarantee the uniqueness of the mediating map -- it rather corresponds
to a condition called quasi-initiality by Wadler
\cite{WadlerRecTypes}. In order to obtain uniqueness, hence to ensure
that $\msf{in^C}$ is an isomorphism, the following implication needs
to be proved for $F$ \cite{Dela13,Keuchel13,Hutton99}.
\begin{gather} \label{church-alg-init1}
 (\forall x: \msf{Fix^C} \ F. \ \ h \ (\msf{in^C} \ F \ x) \ = \ f
\ (\msf{fmap} \ F \ h \ x)) \ \to \ (h \ = \ \msf{fold^C} \ F \ C
\ f) 
\end{gather}
Semantically, the impredicative encoding of the fixed points is
closely associated with a constructor, usually called $\msf{build}$,
that allows for an alternative interpretation of inductive datatypes
in terms of limit constructions, provably equivalent to the initial
algebra semantics \cite{GhaniUV04}.

\subsection{Indexed algebras}

%

A relation can be represented as a function from the type of its
tupled arguments to the type $\msf{P}$ of propositions. From the point
of view of initial semantics, assuming $\msf{P}$ can be represented as
a category, the modular representation of inductively defined
relations only requires a shift of base category. Given a type $K$
(i.e., $K: \msf{Type}$) and assuming it can be represented as a small
category, we can take the category of diagrams of type $K$ in
$\msf{P}$ as the base category for the relations of type $K \to
\msf{P}$. In such category, an endofunctor 
%
$R: \ (K \to \msf{P}) \ \to \ (K \to \msf{P})$
%
that here we call \emph{indexed functor}, is then associated with a
map (\emph{indexed functor map}) that preserves identities and
composition.
\begin{gather}
\begin{array}{lll}
\msf{fmap^I} \ K \ R : \forall \ \{ A \ B : \ K \to \msf{P} \}. 
\ (\forall w:K. \ A \ w \to B \ w) \to (\forall w:K. \ R \ A \ w \to R
\ B \ w)
\end{array}
\end{gather}
From the point of view of the impredicative encoding, 
an $R$-algebra can be characterised as an indexed map, given a carrier
$D: K \to \msf{P}$.
%
\begin{gather}
\msf{Alg^{CI}} \ K \ R \ D \ \defeq \ \forall w:K. \ R \ D \ w \to D \ w
\end{gather}
The corresponding fixpoint operator has type $ ((K \to \msf{P}) \ \to
\ K \to \msf{P}) \to K \to \msf{P} $.
\begin{gather} \msf{Fix^{CI}} \ K \ R \ (w:K) \ 
\defeq \ \forall A: K \to \msf{P}. \ \msf{Alg^{CI}} \ K \ R \ A
\ \to \ A \ w
\end{gather}
The structuring operators can be defined as follows:
\begin{gather} 
\begin{array}{lll}
\msf{fold^{CI}} \ K \ R : \ \forall A \ (f: \msf{Alg^{CI}} \ K \ R \ A)
\ (w: K). \ \msf{Fix^{CI}} \ K \ R \ w \to A \ w \ \defeq 
%
\ \lambda A \ f \ w \ e. \ e \ A \ f
\end{array}
\end{gather}
\begin{gather} 
\begin{array}{lll}
\msf{in^{CI}} \ K \ R \ (w: K) : R \ (\msf{Fix^{CI}} \ K \ R) \ w \to
\msf{Fix^{CI}} \ K \ R \ w \ \defeq \\ 
 \qquad 
\lambda x \ A \ f. \ f \ w \ (\msf{fmap^I} \ K \ R 
\ (\msf{fold^{CI}} \ K \ R \ A \ f) \ w \ x)
\end{array}
\end{gather}
\begin{gather}
\begin{array}{lll}
\msf{out^{CI}} \ K \ R \ (w: K) : \msf{Fix^{CI}} \ K \ R \ w \to R
\ (\msf{Fix^{CI}} \ K \ R) \ w \ \defeq \\ \qquad \msf{fold^{CI}} \ K
\ R \ (R \ (\msf{Fix^{CI}} \ K \ R)) \ (\msf{fmap^{I}} \ K \ R 
\ (\msf{in^{CI}} \ K \ R)) \ w
\end{array}
\end{gather}

\subsection{Proof algebras}   \label{subsection:proof-algebras}

The impredicative encoding makes it comparatively easy to represent
MDTs in Coq, but leaves us with the problem of how to reason
inductively about them. Unlike the in-map of the categorical
semantics, $\msf{in^C}$ is not a constructor -- therefore, structural
induction cannot be applied to a term of type $\msf{Fix^C} \ F$. Let
$P: T \to \msf{P}$ be a property and $T$ the representation of an
inductive datatype in the following goal, which we assume to be
semantically provable by induction on $T$.
\begin{gather} \label{proof-alg-orig1}
\Gamma, w:T \vdash g: P \ w
\end{gather}
However, given $T \defeq \msf{Fix^C} \ F$ and the impredicative
definition of $\msf{Fix^C}$, the type $T$ is not syntactically
inductive, and no conventional induction principle can be
applied. Nevertheless, we can prove
\begin{gather} \label{proof-alg-eq1}
 \forall v:T. \ \exists w:F \ T. \ P \ v \ = \ P \ (\msf{in^C} \ F
 \ w)
\end{gather}
as this follows from the equality $v \ = \ \msf{in^C} \ F
\ (\msf{out^C} \ F \ v)$ which can be proved, provided $\msf{in^C}
\ F$ is shown to be an isomorphism -- e.g., by proving
(\ref{church-alg-init1}). Rewriting (\ref{proof-alg-orig1}) with
(\ref{proof-alg-eq1}), we obtain
\begin{gather}
\Gamma, w:F \ T \vdash g': P \ (\msf{in^C} \ F \ w)
\end{gather}
Here it is possible to apply induction on $w$, since $F \ T$ is an
inductive datatype: however, what we actually get is case analysis --
the recursive arguments in $F \ T$ are hidden in the same sense as
before, as they have type $T$ rather than $F \ T$.

The solution adopted by Delaware \emph{et al.} in \cite{Dela13},
implemented in Coq and supported by MTC/3MT consists of packing an
existential copy of the inductive term together with a proof that it
satisfies the property, using $\Sigma$ types. This involves replacing
the conventional proof with one based on the representation of the
goal as an algebra, i.e., a \emph{proof algebra}.
\begin{gather} \label{proof-alg-alt1}
\Gamma \vdash f: \ \msf{Alg^C} \ F \ (\Sigma v. \ P \ v)
\end{gather}
By folding such an algebra, one obtains
\begin{gather}
\Gamma, w:T \vdash \msf{fold^C} \ F \ (\Sigma v. \ P \ v) \ f \ w:
\ \Sigma v. \ P \ v
\end{gather}
which states something weaker than the original goal
(\ref{proof-alg-orig1}). Nonetheless, under conditions associated with
\emph{well-formed proof algebras} in \cite{Dela13},
(\ref{proof-alg-alt1}) can be strengthened to (\ref{proof-alg-orig1}).
%
This technique is quite general, and it can be applied to inductive
proofs in which the goals may depend on the inductive argument (i.e.,
it can deal with \emph{dependent induction}). However, the proofs that
are obtained in this way are essentially factored into two non-trivial
parts -- the application of a weak induction principle and a
well-formedness proof -- and therefore are quite different from
conventional inductive ones.

\subsection{Looking for a simpler solution}

A natural question arises: is it possible to sacrifice some of the
generality of the MTC approach, to obtain proofs that look more
familiar?
The whole point of using $\Sigma$ types is to hide dependencies: a
solution that does not involve them and so a positive answer to our
question appear more feasible, when we can dispense with the use of
dependent induction, by finding an alternative, equivalent formulation
of the goal. In our schematic example (\ref{proof-alg-orig1}) we get
such reformulation, when we can find $S, \ Q: T \to \msf{P}$ and an
indexed functor $R: (T \to \msf{P}) \to T \to \msf{P} $ such that $S
\defeq \msf{Fix^{CI}} \ T \ R$, the following equivalence holds
\begin{gather} \label{def:agrees}
\mbox{there exists } t \mbox{ s.t. } \ \ \Gamma \vdash t: \ \forall w:
T. \ S \ w \to Q \ w \qquad \mbox{iff \qquad there exists } t' \mbox{
  s.t. } \ \ \Gamma \vdash t': \ \forall w:T. \ P \ w
\end{gather}
and the following is semantically provable, as the new goal, by
induction on $h$:
\begin{gather} \label{def:alter}
 \Gamma, \ w: T, \ h: S \ w \vdash l: Q \ w  \end{gather} 
Intuitively, this means that the dependency of the proof on $w$ can be
lifted to a type dependency, given a sufficiently close analogy
between $T$ as modular inductive datatype and $S$ as modular inductive
predicate, therefore by rather using $h$ of type $S \ w$ as inductive
argument.
Again, we need to expose the inductive structure by shifting to
\begin{gather} 
\label{proof-alg-last}
 \Gamma, \ w: T, \ h: R \ (\msf{Fix^{CI}} \ T \ R) \ w \vdash l': Q
 \ w \end{gather}
and this is not problematic.
However, as before, we end up stuck with case analysis rather than
proper induction. In order to solve this problem, we need to look at
an alternative encoding of fixed points, based on Mendler-style
induction \cite{Mendler91,AbelMU05}. In fact, Mendler's approach makes
it possible to build induction principles into impredicatively encoded
fixed points. Notice that Mendler algebras are used by Delaware
\emph{et al.} \cite{Dela13}, but have a different purpose there (i.e.,
controlling the order of evaluation), from the one we are proposing
here.

\section{Mendler algebras} \label{section:two}

We first present the Mendler-style semantics of inductive datatypes by
introducing Mendler algebras as a category, following Uustalu and Vene
\cite{UustaluV99}. Given a covariant functor $F: \msf{S} \to \msf{S}
$, a Mendler algebra is a pair $\langle C, \Psi \rangle$ where $C:
\msf{S}$ is the carrier and $\Psi \ A: \ (A \to C) \to (F \ A \to C)
$, for each $A: \msf{S}$, is a map from morphisms to morphisms
satisfying $ \Psi \ A \ f \ = \ (\Psi \ C \ \msf{id}_C) \cdot
(\msf{fmap} \ F \ f)$, with $f$ a morphism from $A$ to $C$. A morphism
between Mendler algebras $\langle C_1, \Psi_1 \rangle$ and $\langle
C_2, \Psi_2 \rangle$, is a morphism $h: C_1 \to C_2$ that satisfies $h
\cdot \Psi_1 \ C_1 \ \msf{id}_{C_1} = \Psi_2 \ C_1 \ h$. The Mendler
algebra semantics has been proved equivalent to the conventional one
by Uustalu \emph{et al.}.
Assume $F$ such that the conventional initial $F$-algebra $\langle \mu
F, \msf{in}_F \rangle$ exists. Given the abbreviation
\begin{gather}  \msf{pre\_in}_F \ C \ (m: C \to \mu F) \ \defeq \ \msf{in}_F \cdot
(\msf{fmap} \ F \ m) : \ (F \ C \to \mu F)  \end{gather} 
we can prove the equation 
\begin{gather}  \label{mendler-initial-in}
\msf{in}_F \ = \ \msf{pre\_in}_F \ \mu F \ \msf{id} \end{gather}
%
by the isomorphic character of $\msf{in}_F$. 
The Mendler algebra $\langle \mu F, \ \msf{pre\_in}_F \rangle$ can
thus be shown to be the initial object in its category, and therefore
used as alternative interpretation of the inductive datatype
associated with $F$. For each Mendler algebra $\langle C, \Psi \rangle
$, the unique incoming morphism from the initial Mendler $F$-algebra
can be defined
\begin{gather}  \msf{mfold} \ F \ C \ \Psi \ x \ \defeq \ \Psi
\ (\mu F) \ (\msf{mfold} \ F \ C \ \Psi) \ (\msf{out}_F \ x) \end{gather} 
%
%
Unlike the conventional fixpoint operator, the Mendler one can be
encoded in Coq as an inductive datatype (though using the
impredicative option).
\begin{gather}  
\msf{dt\_def} \ \msf{MFix} \ F \ = \ \msf{pre\_in} \ (C: \msf{S})
\ (b: C \to \msf{MFix} \ F) \ (c: F \ C)
 \end{gather} 
%
However $\msf{in}$, as defined by equation (\ref{mendler-initial-in})
in this setting, is still not a constructor, and the definition of
$\msf{mfold}$ is not structurally recursive. Therefore, also in this
case, it seems more convenient to resort to an impredicative encoding,
following \cite{Mendler91,Dela13}.

\subsection{Impredicative Mendler algebra encoding}

Mendler algebras can be characterised impredicatively by the type of
their structure maps, and a fixpoint operator can be defined as in the
conventional case \cite{Mendler91,Dela13}.
\begin{gather} \qquad \msf{Alg^M} \ F \ C \ \defeq \ \forall A. \ (A \to C) \to (F \ A \to
C) \\
%
\msf{Fix^M} \ F \ \defeq \ \forall C. \ \msf{Alg^M} \ F \ C \to C \end{gather} 
Unlike the conventional case, the type of a Mendler algebra can be
read as specification of an iteration step, where the bound type
variable $A$ represents the type of the recursive calls.  The
corresponding fold operator
\begin{gather} \msf{fold^M} \ F \ C \ f \ x \ \defeq \ x \ C \ f \end{gather} 
indeed has type
\begin{gather}  \msf{fold^M} \ F \ C : \ (\forall A. \ (A \to C) \to (F \ A \to C))
\to (\msf{Fix^M \ F}) \to C \end{gather}
which can represent an induction principle, under the assumption that
the argument to the induction hypothesis is only used therein without
further analysis \cite{Mendler91,AbelMU05}. In-maps and out-maps can
be defined as follows
\begin{gather} 
\begin{array}{lll}
\msf{in^M} \ F \ (x: F (\msf{Fix^M} \ F)): \msf{Fix^M} \ F \ \defeq \ 
%
\lambda  A \ (f: \msf{Alg^M} \ F \ A). \ f
\ (\msf{Fix^M} \ F) \ (\msf{fold^M} \ F \ A \ f) \ x
\end{array}
 \end{gather} 
\begin{gather} 
\begin{array}{lll}
 \msf{out^M} \ F \ (x: \msf{Fix^M} \ F) : F \ (\msf{Fix^M} \ F)
 \ \defeq \ x \ (F \ (\msf{Fix^M} \ F)) \\ \qquad \ (\lambda  A \ (r: A
 \to F \ (\msf{Fix^M} \ F)) \ (a: F \ A). 
 \ \msf{fmap} \ F \ (\lambda y:A. \ \msf{in^M} \ F \ (r \ y)) \ a)
\end{array}
 \end{gather} 
As in the conventional case, impredicative fixpoint definitions give
us quasi-initiality. The uniqueness condition of $\msf{fold^M} \ F \ A
\ f$ that is needed for initiality, in a way which parallels
(\ref{church-alg-init1}), is given by
\begin{gather}  \label{mendler-alg-init1}
\begin{array}{lll}
(\forall x: F \ (\msf{Fix^M} \ F). \ h \ (\msf{in^M} \ F \ x) \ = \ f
  \ (\msf{Fix^M} \ F) \ h \ x) \ \to \ 
h \ = \ \msf{fold^M} \ F \ A \ f
\end{array}
 \end{gather} 
to be proven for a fixed $F$, for every $A: \msf{S}$, $f:\msf{Alg^M}
\ F \ A$ and $h: \msf{Fix^M} \ F \to A$ \cite{UustaluV99}.

\subsection{Indexed Mendler algebras}

As before, we need indexed algebras to deal with relations.  The
definitions are similar to the conventional ones, with $K$ a type, $R:
(K \to \msf{P}) \to (K \to \msf{P}) $ an indexed functor, and $D: K
\to \msf{P}$ an indexed carrier.
\begin{gather}  \msf{Alg^{MI}} \ K \ R \ D \defeq 
\forall A. \ (\forall w:K. \ A \ w \to D \ w) \to \forall w:K. \ R
\ A \ w \to D \ w 
 \end{gather} 
\begin{gather}  \msf{Fix^{MI}} \ K \ R \ w \defeq  
\forall A. \ \msf{Alg^{MI}} \ K \ R \ A \to A \ w
 \end{gather} 
%
%
\begin{gather}  \msf{fold^{MI}} \ K \ R \ D \ 
(f: \msf{Alg^{MI}} \ K \ R \ D) \ (w: K) \ (x: \msf{Fix^{MI}} \ K \ R
  \ w) \defeq x \ D \ f \end{gather}
\begin{gather}  
\begin{array}{lll}
\msf{in^{MI}} \ K \ R \ (w: K) \ (x: R \ (\msf{Fix^{MI}} \ K \ R) \ w):
\msf{Fix^{MI}} \ K \ R \ w \ \defeq \\
\qquad \lambda A \ (f: \msf{Alg^{MI}} \ K \ R \ A). \ f
\ (\msf{Fix^{MI}} \ K \ R) \ (\msf{fold^{MI}} \ K \ R \ A \ f) \ w \ x
\end{array}
 \end{gather} 
\begin{gather} 
\begin{array}{lll}
 \msf{out^{MI}} \ K \ R \ (w: K) \ (x: \msf{Fix^{MI}} \ K \ R \ w) : R
 \ (\msf{Fix^{MI}} \ K \ R) \ w \ = \\
\qquad x \ (R \ (\msf{Fix^{MI}} \ K \ R)) \ (\lambda \ A \ (r: \forall
v. \ A \ v \to R \ (\msf{Fix^{MI}} \ K \ R) \ v) \\ \qquad 
\quad (w: K) \ (a: R \ A \ w). \ \msf{fmap^{I}} \ R \ (\lambda y:A
\ w. \ \msf{in^{MI}} \ K \ R \ w \ (r \ w \ y)) \ a)
\end{array}
\end{gather} 
%
%
%
As an example, we can define inductively a relation $\msf{Eval}:
\ (\msf{Trm} * \msf{Val}) \to \msf{P}$
that agrees with $\msf{eval}$.
\begin{gather}
\begin{array}{lll}
\msf{dt\_def} \ \msf{Eval_G} \ 
(A: \ (\msf{Trm} * \msf{Val}) \to \msf{P})
 \ : \ (\msf{Trm} * \msf{Val}) \to \msf{P} \ = \\
%
 \qquad \msf{ev1}: \ \forall x:\msf{Int}. \ \msf{Eval_G} \ A
\ (\msf{lit} \ x, \msf{val} \ x) \\
 \qquad \msf{ev2}: \ \forall e_1 \ e_2: \msf{Trm}, x_1 \ x_2: \msf{Val}. \ A
 (e_1,x_1) \ \land \ A (e_2,x_2) \to \\
\qquad \qquad \qquad \qquad \qquad \qquad \qquad \qquad \msf{Eval_G} \ A
 \ (\msf{add}(e_1,e_2), \msf{val}((\msf{vv} \ x_1) + (\msf{vv} \ x_2)))
\end{array} \\
\label{def:Eval}
\msf{Eval} \ =_{df} \ \msf{Fix^{MI}} \ (\msf{Trm}*\msf{Val}) \ \msf{Eval_G} 
\end{gather}

\subsection{Proof algebras, Mendler-style}

Reconsider the schematic example in
Section~\ref{subsection:proof-algebras}: the problem in
(\ref{proof-alg-last}) was the missing induction hypothesis, that
cannot be obtained by appealing to the standard inductive principle,
as the recursive occurrences are wrapped in a non-inductive
type. Intuitively, this can be fixed by giving such an hypothesis
explicitly. This would give us a generic representation of the step
lemma in our inductive proof.
%
\begin{gather}  \label{mendler-proof-alg-lemma1}
\begin{array}{lll}
\Gamma, \  h_0 : \forall v:T. \ \msf{Fix^{CI}} \ T \ R \ v \to Q
\ v, \
%
w: T, \ h_1: R \ (\msf{Fix^{CI}} \ T \ R) \ w \ \vdash \ q: Q \ w
\end{array}
 \end{gather} 
However, here the type of $h_0$ is actually too specific to be that of
the induction hypothesis with respect to $h_1$ -- as a result, the
sequent is too weak to take us to the main goal (\ref{def:alter}). At
this point, Mendler's intuition comes into play: under the assumption
that the argument passed to the induction hypothesis is used only
there, without further case analysis, and that therefore we make no
use of its type structure, its type can be represented by a fresh type
variable -- the key feature of Mendler-style induction
\cite{Mendler91,AbelMU05}. We can then strengthen
(\ref{mendler-proof-alg-lemma1}) to the following, more abstract goal.
%
\begin{gather}  \label{mendler-proof-alg-lemma2}
\begin{array}{lll} 
\Gamma, \ A: \msf{Type}, \ h_0 : \forall v:T. \ A \ v \to Q \ v, \ 
%
%
w: T, \ h_1: R \ A \ w \ \vdash \ p: Q \ w
\end{array}
 \end{gather} 
Given $f \defeq \lambda A \ h_0 \ w \ h_1. \ p $, the above is
equivalent to
\begin{gather}  \label{def:mendler-alg}
\Gamma \ \vdash \ f: \ \msf{Alg^{MI}} \ T \ R \ Q  \end{gather} 
Now we have an indexed Mendler algebra. The original goal, equivalent
to (\ref{proof-alg-orig1}) by a reformulation of (\ref{def:agrees})
with $S = \msf{Fix^{MI}} \ T \ R$, can then be obtained by folding,
without need of further adjustments.
\begin{gather}  \Gamma \ \vdash \ \msf{fold^{MI}} \ T \ R \ Q \ f \ : \ \forall
w:T. \ S \ w \to Q \ w \end{gather}
In order to prove (\ref{mendler-proof-alg-lemma2}), case analysis (as
provided in Coq e.g. by \emph{inversion} and \emph{destruct} tactics
\cite{coqart}) can be applied to $h_1$, allowing us to reason on the
structure of $R \ A \ w$. This actually results in doing induction on
that structure, as the induction hypothesis $h_0$ is already there. In
this way, we can minimise the overhead of combining inductive proofs
with modular datatypes. Proving an inductive lemma boils down to
constructing the appropriate Mendler algebra -- the rest is either
conventional, or comes for free. In connection with MDT, such algebras
can be regarded as proof modules, that can be composed together in the
usual sense of case analysis on coproducts \cite{Swier08,Dela13}, in
the same straightforward way as evaluation algebras (the original
motivating example by Swierstra \cite{Swier08}). This sounds
attractive, from the point of view of the applications in which the
relational aspect is predominant, such as structural operational
semantics.

\subsection{Problematic aspects}

Which could be the downsides of the Mendler-based approach?
As already observed, relying on impredicative encodings gives us for
free only a weak semantics of inductive datatypes, i.e., a
quasi-initial one. However, initiality is needed virtually everywhere
in our proofs, to ensure in-maps and out-maps are inverses, i.e.
%
\begin{gather}  (A) \ \ \msf{out^M} \ F \ (\msf{in^M} \ F \ x)  \ = \ x  
\qquad \qquad \qquad
(B) \ \ \msf{in^M} \ F \ (\msf{out^M} \ F \ x) \ = \ x \end{gather}
and similarly for the indexed case. In order to get proper initial
semantics, functor-specific proofs of properties such as
(\ref{mendler-alg-init1}) for base category $\msf{S}$, or the
corresponding one for $K \to \msf{P}$, need to be carried out. This
may be regarded as a general weakness of impredicative approaches
including MTC/3MT \cite{Dela13,Delaware13M}, as remarked by Keuchel and
Schrijvers \cite{Keuchel13}. Nonetheless, in discussing the
well-formedness of Church encodings \cite{Dela13}, Delaware \emph{et
  al.} argue that dealing with this issue is not too hard, as indeed
MTC provides automation for doing so.




A more specific problem is related to the iterative character of
Mendler-style recursion, and correspondingly, to the non-dependent
character of Mendler-style induction. Mendler algebras make it
possible to factor induction into case analysis and folding, but this
restricts induction, in the sense of what is called \emph{Mendler
  iteration} by Abel, Matthes and Uustalu \cite{AbelMU05}: the
argument of the induction hypothesis cannot be used anywhere else,
effectively ruling out dependent induction. This means there are
problems that cannot be solved in their original form. As an example,
MTC \cite{Dela13} proves the type soundness of a language with a
dynamic semantics that is recursively defined as a total evaluation
function. This problem can be reformulated with respect to our
concrete example in Section~\ref{section:zero}, using our definition
of $\msf{eval}$ (\ref{data-rhoX7}).
\begin{gather}  \label{exper1}
\Gamma, e: \msf{Trm}, t: \msf{Typ} \ \vdash \ k : \ \msf{TypOf} \ ( e,
\ t) \to \msf{TypOf} \ (\msf{lit} \circ \msf{vv} \ (\msf{eval} \ e),
\ t) \end{gather}
Using the MTC approach, (\ref{exper1}) can be proved by dependent
induction on the structure of term $e$. Given $\msf{dt\_def}
\ \msf{Typ} \ = \ \msf{N} $ and assuming for simplicity $\msf{TypOf}$
is a conventional inductive predicate
\begin{gather}
\begin{array}{lll}
\msf{dt\_def} & \msf{TypOf} : \ \msf{Trm}*\msf{Typ} \to \msf{P} \ =
\\ & \msf{tof1}: \ \forall v:\msf{Val}. \ \msf{TypOf} \ (\msf{lit}
\circ \msf{vv} \ v, \ \msf{N}) \\
& \msf{tof2}: \ \forall e_1 \ e_2: \msf{Trm}. \ \msf{TypOf}
\ (e_1,\msf{N}) \land \msf{TypOf} \ (e_2,\msf{N}) \to \msf{TypOf}
\ (\msf{add}(e_1,e_2), \msf{N})
\end{array}
\end{gather}
the proof is ultimately based on a proof algebra of type $\msf{Alg^C}
\ \msf{Trm_G} \ (\Sigma e. \ \forall t:\msf{Typ}. \ \msf{TypOf} \ (e,
\ t) \to \msf{TypOf} \ (\msf{lit} \circ \msf{vv} \ (\msf{eval} \ e),
\ t) ) $, although as already noticed, folding this algebra only gives
us the backbone of the whole proof.

This is not possible using our Mendler-style approach, as we cannot
deal with the dependency of the goal on the inductive argument $e$.
What we can do instead, is to rely on the relational formulation of
evaluation given by $\msf{Eval}$ (\ref{def:Eval}), which can be shown
to satisfy (\ref{def:agrees}), 
%
%
and prove
\begin{gather}  \Gamma, e: \msf{Trm}, v: \msf{Val}, \ t: \msf{Typ}, \ h:
\msf{Eval} \ (e,v) \ \vdash \ l : \ \msf{TypOf} \ (e, \ t) \to
\msf{TypOf} \ (\msf{lit} \circ \msf{vv} \ v, \ t)
\end{gather}
reasoning by induction on the structure of $\msf{Eval}$. This
reformulation of the goal essentially matches (\ref{def:alter}). In
this case, a proof can be obtained by simply folding an indexed
Mendler algebra of type $\msf{Alg^{MI}} \ (\msf{Trm}*\msf{Val})
\ \msf{Eval_G} \ (\lambda (e, v). \ \forall t:\msf{Typ}. \ \msf{TypOf}
\ (e, \ t) \to \msf{TypOf} \ (\msf{lit} \circ \msf{vv} \ v, \ t) ) $,
which provides our instance of (\ref{def:mendler-alg}).

An alternative way to obtain a relational equivalent of (\ref{exper1})
is to lift the modular datatype $\msf{Trm}$ to a modular predicate
$\msf{IsTrm}: (\msf{Trm_G} \ \msf{Trm}) \to \msf{P}$, with
$\msf{IsTrm} \ =_{df} \ \msf{Fix^{MI}} \ (\msf{Trm_G} \ \msf{Trm})
\ \msf{IsTrm_G}$, where
\begin{gather} \label{data-rhoXX2}
\begin{array}{lll}
 \msf{dt\_def} \ \msf{IsTrm_G} \ A & = \ \msf{isLit}: \ \forall
 x:\msf{Int}. \ \msf{IsTrm_G} \ A \ (\msf{lit} \ x) \\
& \mid \msf{isAdd}: \ \forall e_1 \ e_2: \msf{Trm}. \ A \ e_1
 \ \land \ A \ e_2 \ \to \ \msf{IsTrm_G} \ A \ (\msf{add} \ (e_1,e_2))
\end{array}
\end{gather}
and then prove 
\begin{gather}  \label{exper11}
\Gamma, e: \msf{Trm}, w: \msf{IsTrm} \ e, t: \msf{Typ} \ \vdash \ k :
\ \msf{TypOf} \ ( e, \ t) \to \msf{TypOf} \ (\msf{lit} \circ \msf{vv}
\ (\msf{eval} \ e), \ t) \end{gather}
reasoning by Mendler induction on $w$.  Notice that $\msf{eval}$ in
the MTC example \cite{Dela13} is actually defined as the fold of a
Mendler algebra, rather than a conventional one, in order to allow for
control over the evaluation order -- this is related to the form of
their semantics though, and completely unrelated to our use of
Mendler-style induction.


\section{Case study}  \label{section:three}


The use of relational formulations appears particularly natural in
specifications based on small-step rules in the style of SOS,
originally introduced by Plotkin \cite{Plotkin04a}. Yet in order to
formulate each relation modularly, we need to build encodings based on
functors that reflect the structure of those relations. This
inevitably makes things more complex, especially when we have to deal
with mutually inductive definitions. In order to test the
applicability of Mendler proof algebras to the formalisation of a
semantic framework, we have formalised a language $\mcal{L}$ with a
comparatively rich syntactic structure, including types ($\msf{Typ}$),
patterns ($\msf{Pat}$), declarations ($\msf{Dec}$) and expressions
($\msf{Exp}$), as well as value environments ($\msf{Env^E}$) and
typing environments ($\msf{Env^T}$). We rely on SOS to give a partial
specification of the language: partial, insofar as we do not specify
any behaviour in case of pattern matching failure -- therefore, we
cannot prove type soundness, which in fact does not hold. However, we
can still prove type preservation -- and this suffices for us, as an
example of the structural complexity we are aiming at.

The full language specification is available with the Coq
formalisation in the companion code at \cite{MACoq}. Here we outline
the specification using conventional dataytpes. The Coq formalisation
is entirely based on modular datatypes, although for simplicity we
rely on monolithic functors (we have not yet implemented the smart
constructor mechanism that facilitates the use of coproducts).
%
\begin{gather}  
\begin{array}{llll}
 \msf{dt\_def} & \msf{Typ} \ = & \msf{ty}(\msf{Id^T}) \mid \msf{Typ}
 \iimp \msf{Typ} \mid \msf{type\_env} (\msf{Env^T}) \\
 \msf{dt\_def} & \msf{Pat} \ = & \msf{vr^p}(\msf{Id},\msf{Typ}) \mid
 \msf{cn^p}(\msf{Id},\msf{Typ}) \mid \msf{apply^p}(\msf{Pat},
 \msf{Pat}) \\
 \msf{dt\_def} & \msf{Dec} \ = & \msf{env}(\msf{Env^E}) \mid
 \msf{match}(\msf{Pat}, \msf{Exp}) \mid \msf{join}(\msf{Dec},
 \msf{Dec}) \\
 \msf{dt\_def} & \msf{Exp} \ = & \msf{vr}(\msf{Id}) \mid
 \msf{cn}(\msf{Id},\msf{Typ}) \mid \msf{closure}(\msf{Env^E}, \msf{Pat},
 \msf{Exp})  \\ & & 
  \mid \msf{apply}(\msf{Exp}, \msf{Exp}) \mid
 \msf{scope}(\msf{Dec}, \msf{Exp}) 
\end{array}
 \end{gather} 
\begin{gather}
\msf{Env} \ A \ \defeq \msf{Id} \to \msf{option} \ A
\qquad \qquad \msf{Env^T} \ \defeq \msf{Env} \ \msf{Typ} 
\qquad \qquad \msf{Env^E} \ \defeq \msf{Env} \ \msf{Exp}
\end{gather} 
The language $\mcal{L}$ is based on simply typed lambda calculus with
pattern matching and first class environments. We use two sets of
identifiers -- $\msf{Id^T}$ for type variables and $\msf{Id}$ for
object variables and constants. Constants and pattern variables are
annotated with types. $\iimp$ is the usual function type
constructor. We use closures instead of lambda abstractions to ensure
values are closed terms and avoid dealing with
substitution. Abstraction is defined over patterns (rather than simply
over variables). Matching patterns with expressions give declarations,
which may evaluate to environments. Declarations can be joined
together and used in scope expressions. Values can be specified as
follows.
\begin{gather}  
\begin{array}{llll}
\mbox{Data values}: \qquad & h \ \in & \ \msf{cn}(x,\tau) \mid
\msf{apply}(h, v) \\
\mbox{Values}: \qquad & v \ \in &
\ \msf{closure}(\rho, p, e) \mid h
\end{array}
 \end{gather} 
The typing relations have the following signatures. Notice that
patterns and values can be typed in a context-free way, unlike
expressions and declarations.
\begin{gather} 
\begin{array}{llll}
\mbox{Patterns}: \qquad & \msf{TypOPat} \ : \ \msf{Pat} * \msf{Typ}
\to \msf{P} \\
\mbox{Environments}: \qquad & \msf{TypOEnv} \ : \ \msf{Env^E} *
\msf{Env^T} \to \msf{P} \\
\mbox{Declarations}: \qquad & \msf{TypODec} \ : \ \msf{Env^T} *
\msf{Dec} * \msf{Typ} \to \msf{P} \\
\mbox{Expressions}: \qquad & \msf{TypOExp} \ : \ \msf{Env^T} *
\msf{Exp} * \msf{Typ} \to \msf{P}
\end{array}
 \end{gather} 
The transition relations have the following signatures.
\begin{gather} 
\begin{array}{llll}
\mbox{Declarations}: \qquad & \msf{DecStep} \ : \ 
\msf{Env^E} * \msf{Dec} * \msf{Dec}  \to \msf{P} \\
\mbox{Expressions}: \qquad & \msf{ExpStep} \ : \ 
\msf{Env^E} * \msf{Exp} * \msf{Exp} \to \msf{P}
\end{array}
 \end{gather} 
Expressions and declarations may depend on each other, and therefore
can only have a mutually inductive definition. Analogously, the
definitions of the typing relations and of the transition relations
for these two syntactic categories involve mutual induction. Therefore
we need to introduce functors to reason about mutually inductively
defined sets, as well as mutually inductively defined relations.

\subsection{Mutually inductive sets}

Two mutually recursive datatypes in the base category $\msf{S}$, can
be represented in terms of bi-functors $F_1, \ F_2: \ \msf{S} *
\msf{S} \to \msf{S}$, where bi-functoriality is expressed as existence
of a map $\msf{fmap^D}$ which satisfies the appropriate form of the
usual preservation properties.
\begin{gather}  
\begin{array}{lll}
\msf{fmap^D} : \ \forall \ \{A_1 \ A_2 \ B_1 \ B_2: \msf{S} \}
\ (f_1: A_1 \to B_1) \ (f_2: A_2 \to B_2). \
%
F \ (A_1, A_2) \to F \ (B_1, B_2)
\end{array} \\
%
\begin{array}{lll}
\msf{fmap^D} \ g_1 \ g_2 \ (\msf{fmap^D} \ f_1 \ f_2) \ = \ 
\msf{fmap^D} \ (g_1 \cdot f_1) \ (g_2 \cdot f_2) \\
\msf{fmap^D} \ \msf{id}_A \ \msf{id}_B \ = \ \msf{id}_{F A B}
\end{array}
 \end{gather} 
The definitions of Mendler bi-algebra, fixpoint and fold operators can
be given using pairs.
\begin{gather}  
\begin{array}{lll}
\msf{Alg^{D}} \ (F_1, F_2) \ (C_1, C_2) \ \defeq \  
( \forall A_1 \ A_2. \ (A_1 \to C_1) \to (A_2 \to C_2) \to F_1
\ (A_1, A_2) \to C_1,\\ 
\qquad \qquad \qquad \qquad \qquad \qquad \ \forall A_1
\ A_2. \ (A_1 \to C_1) \to (A_2 \to C_2) \to F_2 \ (A_1, A_2) \to C_2)
\end{array}
 \end{gather} 
\begin{gather}  
\begin{array}{lll}
\msf{Fix^D} \ (F_1, F_2) \ \defeq \ (\forall A_1 \ A_2. \ \msf{Alg^D}
\ (F_1, F_2) \ (A_1, A_2) \to A_1,  \\ 
\qquad \qquad \quad \qquad \qquad 
\forall A_1 \ A_2. \ \msf{Alg^D} \ (F_1, F_2) \ (A_1, A_2) \to A_2) 
\end{array}
\end{gather} 
\begin{gather} 
\begin{array}{lll}
\msf{fold^D_1} \ (F_1, F_2) \ (C_1, C_2) \ (f: \msf{Alg^D} \ (F_1,
F_2) \ (C_1, C_2)): \\ 
\qquad \qquad \qquad 
\msf{fst} \ (\msf{Fix^D} \ (F_1, F_2)) \to C_1
\ \defeq \ \lambda e. \ e \ C_1 \ C_2 \ f
\end{array}
\end{gather} 
\begin{gather} 
\begin{array}{lll}
\msf{fold^D_2} \ (F_1, F_2) \ (C_1, C_2) \ (f: \msf{Alg^D} \ (F_1,
F_2) \ (C_1, C_2)): \\
\qquad \qquad \qquad \msf{snd} \ (\msf{Fix^D} \ (F_1, F_2)) \to C_2
\ \defeq \ \lambda e. \ e \ C_1 \ C_2 \ f
\end{array}
\end{gather} 
All the syntactic categories of $\mcal{L}$ can then be represented as
MDTs, using bi-functors for mutually defined $\msf{Decl}$ and
$\msf{Exp}$.
\begin{gather}  
\begin{array}{llll}
 \msf{dt\_def} \ \msf{Typ_G} \ T \ = \ \msf{ty}(\msf{Id^T}) \mid
 T \iimp T \mid \msf{type\_env} \ (\msf{Env^T} \ T) 
\qquad \qquad \qquad
\msf{Typ} \ =_{df} \ \msf{Fix^M} \ \msf{Typ_G}
\\
 \msf{dt\_def} \ \msf{Pat_G} \ P \ = \ \msf{vr^p}(\msf{Id},T) \mid
 \msf{cn^p}(\msf{Id},T) \mid \msf{apply^p}(P,P) 
\qquad \qquad \qquad \quad 
\msf{Pat} \ =_{df} \ \msf{Fix^M} \ \msf{Pat_G}
\\
 \msf{dt\_def} \ \msf{Dec_G} \ D \ E \ = \ \msf{env}(\msf{Env} \ E) \mid
 \msf{match}(\msf{Pat}, E) \mid \msf{join}(D,D) \\
 \msf{dt\_def} \ \msf{Exp_G} \ D \ E \ = \ \msf{vr}(\msf{Id}) \mid
 \msf{cn}(\msf{Id},\msf{Typ}) \mid \msf{closure}(\msf{Env} \ E, \msf{Pat},
 E) \  \mid \msf{apply}(E, E) \mid
 \msf{scope}(D, E) \\
 \qquad \msf{Dec} \ =_{df} \ \msf{fst} \ (\msf{Fix^D}
 \ (\msf{Dec_G},\msf{Exp_G})) 
\qquad \qquad \msf{Exp} \ =_{df} \ \msf{snd} \ (\msf{Fix^D}
\ (\msf{Dec_G},\msf{Exp_G}))
\end{array}
 \end{gather}

\subsection{Mutually inductive relations}

Given types $K_1, K_2$, two mutually recursive relations depending on
such types in base categories $K_1 \to \msf{P}$, $K_2 \to \msf{P}$,
can be represented by indexed bi-functors $R_1, R_2$, with
\begin{gather} 
\begin{array}{lll}
R_1 \ K_1: (K_1 \to \msf{P}) * (K_2 \to \msf{P}) \to (K_1 \to
\msf{P}) \qquad \quad  
R_2 \ K_1: (K_1 \to \msf{P}) * (K_2 \to \msf{P}) \to (K_2 \to \msf{P})
\end{array}
\end{gather} 
characterised by maps
\begin{gather}  
\begin{array}{lll}
\msf{fmap^H_1} \ (K_1, K_2) \ R_1: \ \forall \ \{ A_1 \ A_2 : \ K_1
\to \msf{P} \} \ \{B_1 \ B_2 : \ K_2 \to \msf{P} \}.  \\
\qquad (\forall w:K_1. \ A_1 \ w \to B_1 \ w) \to (\forall
w:K_2. \ A_2 \ w \to B_2 \ w) \to \\
\qquad \qquad 
\forall w:K_1. \ R_1 \ (A_1, A_2) \ w \to R_1 \ (B_1, B_2) \ w
\end{array} \\
%
\begin{array}{lll}
\msf{fmap^H_2} \ (K_1, K_2) \ R_2: \ \forall \ \{ A_1 \ A_2 : \ K_1
\to \msf{P} \} \ \{B_1 \ B_2 : \ K_2 \to \msf{P} \}.  \\
\qquad (\forall w:K_1. \ A_1 \ w \to B_1 \ w) \to (\forall
w:K_2. \ A_2 \ w \to B_2 \ w) \to \\
\qquad \qquad
 \forall w:K_2. \ R_2 \ (A_1, A_2) \ w \to R_2 \ (B_1, B_2) \ w
\end{array}
\end{gather} 

\noindent Given carriers $D_1: K_1 \to \msf{P}$, $D_2:K_2 \to
\msf{P}$, we can now define indexed Mendler bi-algebras and the
associated notions (see \cite{MACoq} for more details).
\begin{gather}  
\begin{array}{lll}
\msf{Alg^H} \ (K_1, K_2) \ (R_1, R_2) \ (D_1, D_2) \ \defeq \\
\qquad  ( \forall
A_1 \ A_2. \ (\forall w:K_1. \ A_1 \ w \to D_1 \ w) \to (\forall
w:K_2. \ A_2 \ w \to D_2 \ w) \to \\
\qquad \qquad 
 \forall w: K_1. \ R_1 \ (A_1, A_2)
\ w \to D_1 \ w,\\
 \qquad \forall A_1
\ A_2. \ (\forall w:K_1. \ A_1 \ w \to D_1 \ w) \to (\forall
w:K_2. \ A_2 \ w \to D_2 \ w) \to \\
\qquad \qquad \forall w: K_2. \ R_2 \ (A_1, A_2)
\ w \to D_2 \ w )
\end{array}
\end{gather} 
%
%
\begin{gather}  
\begin{array}{lll}
\msf{Fix^H} \ (K_1, K_2) \ (R_1, R_2) \ \defeq \\
\qquad (\lambda w:K_1. \ \forall
A_1 \ A_2. \ \msf{Alg^H} \ (K_1,K_2) \ (R_1, R_2) \ (A_1, A_2) \to
A_1 \ w, \\
\qquad \lambda w:K_2. \ \forall A_1
\ A_2. \ \msf{Alg^H} \ (K_1,K_2) \ (R_1, R_2) \ (A_1, A_2) \to A_2
\ w)
\end{array}
\end{gather} 
%

\begin{gather} 
\begin{array}{lll}
\msf{fold^H_1} \ (K_1,K_2) \ (R_1, R_2) \ (D_1, D_2) \ (f:
\msf{Alg^H} \ (K_1,K_2) \ (R_1, R_2) \ (D_1, D_2)) \ (w:K_1):
\\ \qquad \msf{fst} \ (\msf{Fix^H} \ (K_1,K_2) \ (R_1, R_2)) \ w \to
D_1 \ w \ \defeq \lambda w \ e. \ e \ D_1 \ D_2 \ f
\end{array}
\end{gather} 
\begin{gather} 
\begin{array}{lll}
\msf{fold^H_2} \ (K_1,K_2) \ (R_1, R_2) \ (D_1, D_2) \ (f:
\msf{Alg^H} \ (K_1,K_2) \ (R_1, R_2) \ (D_1, D_2)) \ (w:K_2):
\\ \qquad \msf{snd} \ (\msf{Fix^H} \ (K_1,K_2) \ (R_1, R_2)) \ w \to
D_2 \ w \ \defeq \lambda w \ e. \ e \ D_1 \ D_2 \ f
\end{array}
\end{gather} 
While the typing relations for patterns $\msf{TypOPat}$ can be
represented modularly using an indexed functor and $\msf{Fix^I}$, the
corresponding relations for declarations and expressions,
i.e. $\msf{TypODec}$ and $\msf{TypOExp}$ respectively, are mutually
defined and therefore need to be represented as indexed bi-functors
closed by $\msf{Fix^H}$. Such is also the case for $\msf{DecStep}$ and
$\msf{ExpStep}$, which can be defined as follows, given the
corresponding indexed bi-functors $\msf{DecStep_G}: \ (\msf{Env^E} *
\msf{Dec} * \msf{Dec} \to \msf{P}, \ \msf{Env^E} * \msf{Exp} *
\msf{Exp} \to \msf{P}) \to \msf{Env^E} * \msf{Dec} * \msf{Dec} \to
\msf{P}$, and $\msf{ExpStep_G}: \ (\msf{Env^E} * \msf{Dec} * \msf{Dec}
\to \msf{P}, \ \msf{Env^E} * \msf{Exp} * \msf{Exp} \to \msf{P}) \to
\msf{Env^E} * \msf{Exp} * \msf{Exp} \to \msf{P}$.
\begin{gather} 
\msf{DecStep} \ =_{df} \ \msf{fst} \ (\msf{Fix^H} \ (\msf{Env^E} *
\msf{Dec} * \msf{Dec}, \ \msf{Env^E} * \msf{Exp} * \msf{Exp})
\ (\msf{DecStep_G}, \ \msf{ExpStep_G})) \\
\msf{ExpStep} \ =_{df} \ \msf{snd} \ (\msf{Fix^H} \ (\msf{Env^E} *
\msf{Dec} * \msf{Dec}, \ \msf{Env^E} * \msf{Exp} * \msf{Exp})
\ (\msf{DecStep_G}, \ \msf{ExpStep_G}))
\end{gather}

\subsection{Type preservation}

Type preservation in $\mcal{L}$ can be expressed as follows
\begin{gather}  \label{main}
\begin{array}{lll}
\Gamma, \rho: \msf{Env^E} \ \vdash \ 
(\forall (d_1 \ d_2: \msf{Dec}). \ \msf{DecStep} \ (\rho,d_1,
d_2) \to \msf{DecTSafe} \ (\rho, d_1,d_2) \\
\qquad \qquad \ \ \land \ (\forall (e_1 \ e_2:
\msf{Exp}). \ \msf{ExpStep} \ (\rho,e_1, e_2) \to
\msf{ExpTSafe} \ (\rho, e_1,e_2)
\end{array}
 \end{gather} 
where
\begin{gather} 
\begin{array}{lll}
\msf{DecTSafe} \ (\rho, d_1,d_2) \defeq \ \forall (t: \msf{Typ})
\ (\gamma:\msf{Env^T}). \\
\qquad \qquad \msf{TypOEnv} \ (\rho,\gamma) \to
\msf{TypODec} \ (\gamma,d_1, t) \to \msf{TypODec}
\ (\gamma, d_2, t) \\
\msf{ExpTSafe} \ (\rho, e_1,e_2) \defeq \ \forall (t: \msf{Typ})
\ (\gamma:\msf{Env^T}). \\
\qquad \qquad \msf{TypOEnv} \ (\rho,\gamma) \to
\msf{TypOExp} \ (\gamma,e_1, t) \to \msf{TypOExp}
\ (\gamma, e_2, t)
\end{array}
 \end{gather} 
The context $\Gamma$ includes premises of shape
\begin{gather}
 (\mbox{\it IN} \ x \ = \ \mbox{\it IN} \ y) \ \to \ (x \ = \ y)
\end{gather}
where $\mbox{\it IN} \ $ is the in-map for one of the datatypes --
such premises can be discharged when the corresponding initiality
conditions (\ref{mendler-alg-init1}) are proven. It also includes
premises of shape
\begin{gather}
\forall x:D_G, IsD_G \ x.
\end{gather}
where $D_G$ is the unfolding of a modular datatype $D$, and $IsD_G$ is
the unfolding of a modular predicate $IsD$ that represents the
relational lifting of $D$, in the sense of our example
(\ref{data-rhoXX2}). Such premises are needed, as the proof involves
sublemmas that are proved by induction on the syntactic categories --
and so, for instance, $\msf{Typ_G} \ \msf{Typ}$ has to be lifted to
$\msf{IsTyp_G}: (\msf{Typ_G} \ \msf{Typ} \to \msf{P}) \to \msf{Typ_G}
\ \msf{Typ} \to \msf{P}$.

Crucially, the pair of $\msf{DecTSafe}$ and $\msf{ExpTSafe}$ can be a
carrier for the indexed bi-functor determined by $\msf{DecStep}$ and
$\msf{ExpStep}$. In order to prove type preservation by mutual
induction on the structure of $\msf{DecStep}$ and $\msf{ExpStep}$, we
define an indexed Mendler bi-algebra that has $(\msf{DecTSafe},
\msf{ExpTSafe})$ as indexed carrier, where the index types are
$\msf{Env^E} * \msf{Dec} * \msf{Dec}$ and $\msf{Env^E} * \msf{Exp} *
\msf{Exp}$
\begin{gather}  
\begin{array}{lll}
\msf{TPAlg} \ \defeq \ \msf{Alg^H} \ (\msf{Env^E} * \msf{Dec} *
\msf{Dec}, \ \msf{Env^E} * \msf{Exp} * \msf{Exp}) \\ \qquad \qquad
\qquad \qquad (\msf{DecStep_G}, \ \msf{ExpStep_G}) \ (\msf{DecTSafe},
\ \msf{ExpTSafe})
\end{array}
 \end{gather}  
After finding proofs $ f_1 : \msf{fst} \ \msf{TPAlg} $ and $ f_2 :
\msf{snd} \ \msf{TPAlg} $, we can construct a proof of (\ref{main}) by
applying to them $\msf{fold^H_1}$ and $\msf{fold^H_2}$, respectively
(see \cite{MACoq} for details).

\section{Conclusion}

Motivated by the importance of modularity in program development,
semantics and verification, we have discussed the use of MDTs, their
semantic foundations and their impredicative encoding along the lines
of existing work \cite{Dela13,Keuchel13,Swier08}. We have shown how
impredicative MDT encodings based on Mendler algebras can be used to
reason about inductively defined relations, in a way that is
comparatively close to a more conventional style of reasoning based on
closed datatypes, by providing a simpler notion of proof algebra, if
less general, than the one proposed by Delaware \emph{et al.}
\cite{Dela13}. Our approach can be regarded as a novel application of
Mendler-style induction \cite{Mendler91,AbelMU05,UustaluV99}, as well
as a technique that could be integrated in existing frameworks based
on the impredicative encoding, such as MTC/3MT 
\cite{Dela13,Delaware13M}. Mendler's original insight \cite{Mendler91}
was in the semantics of inductive datatypes -- the case made here, is
for using that insight as a modular proof technique. From the point of
view of possible applications to semantics and verification in
frameworks such as OTT \cite{OTT}, the relational style that can be
supported seems to fit in well with SOS and in particular with
component-based approaches, such as the one proposed by Churchill,
Mosses, Sculthorpe and Torrini \cite{TAOSD}. Our plans for future work
include integrating our technique in MTC/3MT, and comparing this
approach with the container-based one proposed by Keuchel and
Schrijvers \cite{Keuchel13}.

\paragraph{Acknowledgments:}
We thank Steven Keuchel, Neil Sculthorpe, Casper Bach Poulsen and the
anonymous reviewers for feedback on earlier versions, and members of
the Theory Group at Swansea University, including Peter Mosses, Anton
Setzer and Ulrich Berger, for discussion. The writing of this paper
has been supported by EU funding (H2020 FET) to KU Leuven for the
\textsc{GRACeFUL} project. Preliminary work was funded by the EPSRC
grant (EP/I032495/1) to Swansea University for the \textsc{PLanCompS}
project.




\bibliographystyle{eptcs}

\bibliography{fics}

\end{document}

%% file: lightmacros2.tex
\newcommand{\escmathcom}[3]{ \newcommand{#1}[#2]{\mbox{#3}}}
\escmathcom{\mcal}{1}{$\mathcal{#1}$}
\escmathcom{\msf}{1}{$\mathsf{#1}$}
\escmathcom{\lrelll}{1}{\xleftarrow{#1}}

\escmathcom{\iimp}{0}{$\Rightarrow$}

\escmathcom{\dv}{0}{$\mbox{\sf dv}$}
\escmathcom{\pt}{0}{$\mbox{\sf pt}$}
\escmathcom{\Tsort}{0}{$\mathbf{T}$}
\escmathcom{\Psort}{0}{$\mathbf{P}$}
\escmathcom{\FV}{0}{$\mbox{\sf fv}$}
\escmathcom{\floc}{0}{$\mbox{\sf flv}$}
\escmathcom{\Gfloc}{0}{$\mbox{\sf gflv}$}
\escmathcom{\mres}{0}{$\mbox{\sf mres}$}
\escmathcom{\Enc}{0}{$\mbox{\sf enc}$}
\escmathcom{\lnand}{0}{$\&$}
\escmathcom{\lnforall}{0}{$\forall$}
\escmathcom{\injforall}{0}{$\hat{\forall}$}
\escmathcom{\dynex}{0}{$\hat{\exists}$}
\escmathcom{\dynall}{0}{$\hat{\forall}$}
\escmathcom{\lnlambda}{0}{$\hat{\lambda}$}
\escmathcom{\lnLambda}{0}{$\hat{\Lambda}$}
\escmathcom{\caret}{0}{\^{}}
\escmathcom{\nil}{0}{$\msf{nil}$}
\escmathcom{\dbar}{1}{$\bar{#1}$}
\escmathcom{\lsem}{0}{$\llbracket$}
\escmathcom{\rsem}{0}{$\rrbracket$}
\escmathcom{\lnvareps}{0}{$\hat{\xi}$}
\escmathcom{\vareps}{0}{$\xi$}
\escmathcom{\One}{0}{$\bf{1}$}
\escmathcom{\Kappa}{0}{$\Box$}
\escmathcom{\Alpha}{0}{$\Theta$}
\escmathcom{\upnu}{0}{$\nu$}
\escmathcom{\lequiv}{0}{$\hat{\equiv}$}
\escmathcom{\downmapsto}{0}{$\downharpoonright$}
\escmathcom{\ddown}{0}{$\downharpoonleft$}
\escmathcom{\uup}{0}{$\upharpoonright$}
\escmathcom{\szero}{0}{$$}
\escmathcom{\dzero}{0}{$\cdot$}

\escmathcom{\LocS}{0}{$\msf{Loc}$}
\escmathcom{\LinS}{0}{$\msf{Lin}$}
\escmathcom{\GlobS}{0}{$\msf{Glb}$}

\escmathcom{\DD}{0}{$ \ | \ $}
\escmathcom{\IN}{0}{$ \ + \ $}

\escmathcom{\U}{1}{$U({#1})$}

\escmathcom{\Loc}{0}{$\Xi$}
\escmathcom{\Spat}{0}{$\Phi$}
\escmathcom{\Glob}{0}{$\Gamma$}
\escmathcom{\Cont}{0}{$\Delta$}

\escmathcom{\upmapsto}{0}{$\upharpoonright$}
\escmathcom{\plus}{0}{$\&$}
\escmathcom{\defeq}{0}{$\ =_{df} \ $}
\escmathcom{\ineg}{0}{$\neg$}
\escmathcom{\grax}{0}{$\vDash$}
\escmathcom{\VDASH}{0}{$\vdash$}
\escmathcom{\VDASHS}{0}{$ \ \vdash \ $}
\escmathcom{\MID}{0}{$ \ \mid \ $}

\escmathcom{\at}{0}{$@$}

\escmathcom{\edownarrow}{0}{$\downarrow^e$}

\escmathcom{\redeqA}{3}{$ {#1} \equiv {#2} [{#3}] $}
\escmathcom{\redeqB}{3}{$ {#1} \equiv_{[{#3}]} {#2} $}

\escmathcom{\subst}{3}{$ {#1} [{#2} / {#3}] $}
\escmathcom{\rw}{2}{$ {#1} [{#2}] $}

\escmathcom{\copie}{0}{$ \msf{copy} $}
\escmathcom{\an}{0}{$ \msf{at} $}
\escmathcom{\twice}{2}{$ \msf{pair}({#1},{#2}) $}
\escmathcom{\alloc}{0}{$ \msf{alloc} $}

\escmathcom{\redux}{0}{$ \ \longrightarrow \ $}
\escmathcom{\eredux}{0}{$ \ \longrightarrow_e \ $}
\escmathcom{\ereduX}{0}{$ \ \longrightarrow_e^* \ $}
\escmathcom{\aredux}{0}{$ \ \longrightarrow_A \ $}
\escmathcom{\bredux}{0}{$ \ \longrightarrow_{B} \ $}
\escmathcom{\reduxT}{0}{$ \ \longrightarrow^* \ $}
\escmathcom{\reduX}{0}{$ \ \longrightarrow^* \ $}
\escmathcom{\areduX}{0}{$ \ \longrightarrow_{A}^* \ $}
\escmathcom{\redequiv}{0}{$ \ \equiv \ $}

\escmathcom{\tranred}{0}{$ \ \rightarrow_{tr} \ $}
\escmathcom{\instred}{0}{$ \ \rightarrow_{ir} \ $}
\escmathcom{\evalred}{0}{$ \ \rightarrow_{er} \ $}
\escmathcom{\tranext}{0}{$ \ \rightarrow_{te} \ $}
\escmathcom{\instext}{0}{$ \ \rightarrow_{ie} \ $}
\escmathcom{\somered}{0}{$ \ \twoheadrightarrow_{r} \ $}
\escmathcom{\patred}{0}{$ \ \rightarrow_{ptr} \ $}
\escmathcom{\paired}{0}{$ \ \rightarrow_{pir} \ $}

\escmathcom{\avdash}{1}{$ \vdash^{#1} $}
\escmathcom{\bvdash}{1}{$ \vdash {#1} \into $}
\escmathcom{\vvdash}{1}{$ \into {#1} \vdash $}

\escmathcom{\lletexp}{4}{$\msf{let} \ {#1} \leftarrow^l  
                                        {#2} \ \msf{in} \ {#3}$}

\escmathcom{\letexp}{4}{$\msf{let} \ {#1} \overset{ \ #4}{\leftarrow}  
                                        {#2} \ \msf{in} \ {#3}$}

\escmathcom{\letexpX}{3}{$\msf{let} \ {#1} \mathrel{\mathop{\leftarrow}^{\mathrm{l}}}  
                                        {#2} \ \msf{in} \ {#3}$}
\escmathcom{\letex}{3}{$\msf{let} \ {#1}={#2} \ \msf{in} \ {#3}$}

\escmathcom{\lock}{2}{$\langle {#1} \dagger {#2} \rangle$}
\escmathcom{\fresh}{2}{$\langle {#1} \# {#2} \rangle $}
\escmathcom{\export}{2}{$\langle {#1} \triangleleft {#2} \rangle $}
\escmathcom{\scope}{2}{$\langle {#1} ; {#2} \rangle $}

\escmathcom{\encl}{2}{$\langle {#2} \ddown {#1} \rangle$}
\escmathcom{\pair}{2}{$\langle {#2} ; {#1} \rangle $}
\escmathcom{\sg}{1}{$  \# {#1}  $}
\escmathcom{\patt}{2}{$ [ {#1} \leftarrow {#2}] $}

\escmathcom{\typ}{1}{$ \msf{types}({#1}) $}
\escmathcom{\vars}{1}{$ \msf{vars}({#1}) $}

\escmathcom{\locExCons}{3}{$\langle {#1} \ddown {#2} \at \ {#3} \rangle$}
\escmathcom{\locPattOld}{2}{$\langle {#1} \ddown {#2} \rangle$}
\escmathcom{\locPatt}{2}{$ {#1} \ddown {#2} $}
\escmathcom{\lDec}{2}{$\langle {#2} \ddown {#1} \rangle$}
\escmathcom{\locAllDes}{2}{$\langle {#1} \# {#2} \rangle$}
\escmathcom{\locDes}{2}{$\langle {#1} \# {#2} \rangle $}

\escmathcom{\into}{0}{$\ \triangleleft \ $}
\escmathcom{\ENV}{0}{$\ \triangleright \ $}
\escmathcom{\add}{1}{$\ \_ + [{#1}] \ $}

\newcounter{observation}
\renewcommand{\theobservation}
{\arabic{observation}}

{\end{description}}

\newcounter{prop}
\renewcommand{\theprop}
{\arabic{prop}}

{\end{description}}